\documentclass[11pt,a4paper]{article}
\usepackage{jcappub}
\usepackage{mathrsfs}
\title{Cosmic histories of star formation and reionization: An analysis with a power-law approximation}

\author[a]{Yun-Wei Yu,}\emailAdd{yuyw@phy.ccnu.edu.cn}

\author[b]{K. S. Cheng,}%

\author[c]{M. C. Chu,}

\author[c]{S. Yeung}

\affiliation[a]{Institute of Astrophysics, Central China Normal
University, 152 Luoyu Road, Wuhan 430079, China}

\affiliation[b]{Department of Physics, The University of Hong Kong,
Pokfulam Road, Hong Kong, China}

\affiliation[c]{Department of Physics and Institute of Theoretical
Physics, The Chinese University of Hong Kong, Shatin, New
Territories, Hong Kong, China}

\abstract{With a simple power-law approximation of high-redshift
($\gtrsim3.5$) star formation history, i.e., $\dot{\rho}_*(z)\propto
[(1+z)/4.5]^{-\alpha}$, we investigate the reionization of
intergalactic medium (IGM) and the consequent Thomson scattering
optical depth for cosmic microwave background (CMB) photons. A
constraint on the evolution index $\alpha$ is derived from the CMB
optical depth measured by the {\it Wilkinson Microwave Anisotropy
Probe} (WMAP) experiment, which reads
$\alpha\approx2.18\lg{\mathscr{N}_{\gamma}}-3.89$, where the free
parameter $\mathscr{N}_\gamma$ is the number of the escaped ionizing
ultraviolet photons per baryon. At the same time, the redshift $z_f$
at which the IGM is fully ionized can also be expressed as a
function of $\alpha$ as well as $\mathscr{N}_{\gamma}$. By further
taking into account the implication of the Gunn-Peterson trough
observations to quasars for the full reionization redshift, i.e.,
$6\lesssim z_f \lesssim7$, we obtain $0.3\lesssim\alpha\lesssim1.3$
and $80\lesssim\mathscr{N}_{\gamma}\lesssim230$. For a typical
number of $\sim4000$ of ionizing photons released per baryon of
normal stars, the fraction of these photons escaping from the stars,
$f_{\rm esc}$, can be constrained to within the range of
$(2.0-5.8)\%$.}


\begin{document}

\maketitle

\flushbottom

\section{Introduction}
The cosmic history of star formation (especially at high redshifts)
is one of long outstanding problems in astrophysical cosmology,
which is being probed by a growing number of observations such as
through Lyman break galaxies (LBG) \cite{bouw08,bouw11}, Ly-$\alpha$
emitter galaixes \cite{ota08}, and gamma-ray bursts (GRBs)
\cite{kist09,wang09,wang11,robe12,ishi11}. One of the most important
consequences of star formation is the reionization of intergalactic
medium (IGM) that has been recombined at redshift $z\sim 1100$,
although some other sources (e.g., accreting massive black holes)
may also provide contributions to the reionization. The answer to
the question that how and when IGM was reionized must hold many
clues to the nature of the first generation of light sources and
also to the onset of structure formation in cold dark matter
cosmologies. During reionization, the librated free electrons can
scatter cosmic microwave background (CMB) photons, resulting in an
increase in the degree of the anisotropies of both CMB temperature
and polarization \cite{haim99}.

Therefore, since the launch of the {\it Wilkinson Microwave
Anisotropy Probe} ({\it WMAP}) satellite, it has become of great
interest to derive constraints on the cosmic histories of star
formation and reionization from CMB observations. As one of the most
direct results, the {\it WMAP-7} observation infers a Thomson
scattering optical depth of CMB photons $\tau=0.088\pm 0.015$
\cite{lars11,koma11}. Then, the redshift for full reionization can
easily be derived to $z_f =10.5\pm1.2$ with the assumption that the
transition from a fully neutral to a fully ionized IGM is abrupt.
However, deeper implications of the optical depth always invoke an
elaborate description for the star formation history, either by
theoretical modelings in the hierarchical formation model
\cite{haim03a,shul08} or by observational determinations from some
specific observations \cite{wyit10}. In this paper, we alternatively
take a power-law assumption for the high-redshift star formation
history as an effective description, where all of the relevant
physical details are packed into a free parameter (i.e., the power
law index). The power-law behavior could be regarded as the first
order approximation of the realistic history, which in principle
allows us to get some general and substantial inferences.

\section{Model}
First of all, all analyses throughout the paper is based on the
assumption that cosmic reionization is dominated by the early
generation of stars. The rate of ionizing ultraviolet photons
escaping from stars into IGM reads
\begin{eqnarray}
\dot{n}_{\gamma}(z)={\dot{\rho}_*(z)\over m_B}N_{\gamma}f_{\rm
esc},\label{phnum}
\end{eqnarray}
where $\dot{\rho}_*(z)$ is the star formation rate (SFR), $m_B$ is
the mass of a baryon, $N_{\gamma}$ is the number of ionizing
ultraviolet photons released per baryon of the stars, and $f_{\rm
esc}$ is the fraction of these photons escaping from the stars. In
literature, the value of $N_{\gamma}$ is usually taken to be $\sim
4000$ for a Salpeter stellar initial mass function and a metallicity
$\sim 0.05 Z_{\odot}$ (e.g., \cite{bark01}). For the value of
$f_{\rm esc}$, a wild range from a few per cent to $\sim 20$ per
cent was suggested by some observations
\cite{stei01,fern03,shap06,sian07}. Similar values were also
predicted by some theoretical modelings and simulations. For
example, Razoumov \& Sommer-Larsen \cite{razo06} found that $f_{\rm
esc}$ evolves from $\sim1-2$ percent at $z=2.39$ to $\sim6-10$
percent at $z=3.6$. In this paper, we would use the product of
$N_{\gamma}$ and $f_{\rm esc}$ as a single parameter denoted by
$\mathscr{N}_{\gamma}\equiv N_{\gamma}f_{\rm esc}$, which is
considered to be completely free. Accompanying with star formation,
IGM was reionized. The density of the liberated electrons $n_e$ can
be connected to the hydrogen density $n_H$ by $n_e=(1+y)n_Hx$, where
$x=n_{\rm H~II}/n_{\rm H}$ represents the fraction of ionized
hydrogen and $y$ is introduced by considering the ionization of
helium. By assuming that the helium was only once ionized, we have
$y=Y/(4X)\approx 0.08$, where $X=0.7523$ and $Y=0.2477$ are the mass
fractions of the hydrogen and helium, respectively, at the
reionization era \cite{peim07}. Following ref. \cite{bark01}, the
evolution of $x$ can be determined by
\begin{eqnarray}
{dx\over dz}=\left[{\dot{n}_{\gamma}\over (1+y)n^0_{\rm
H}}-\alpha_B{C}(1+z)^3(1+y)n^0_{\rm H}x\right]{dt\over
dz}\label{xz},
\end{eqnarray}
where $n^0_{\rm H}=1.9\times10^{-7}~\rm cm^{-3}$ is the present
number density of hydrogen, $\alpha_B=2.6\times10^{-13}\rm
~cm^3s^{-1}$ is the recombination coefficient of netral hydrogen for
electron temperature of $\sim10^4$ K, $C\equiv {\langle n_{\rm
H~_{II}}^2\rangle / \langle n_{\rm H~_{II}}\rangle^2}$ is the
clumping factor of ionized gas, and $dt/dz=-1/[H(z)(1+z)]$ with
$H(z)=H_0\sqrt{\Omega_m(1+z)^3+\Omega_{\Lambda}}$,
$H_0=71\rm~km~s^{-1}Mpc^{-1}$, $\Omega_m=0.267$, and
$\Omega_{\Lambda}=0.734$ \cite{lars11}.

For a given reionization history $x(z)$, the CMB optical depth can
be calculated by integrating the electron density times the Thomson
cross section along proper length
\begin{eqnarray}
{\tau}&=&\int \sigma_Tn_e dl\nonumber\\
&=&-(1+y)\sigma_Tn_{\rm H}^0c\int_0^{z_{h}} (1+z)^3x(z){dt\over
dz}dz.\label{tau1}
\end{eqnarray}
The {\it WMAP-7} observation gave $\tau=0.088\pm 0.015$, which hence
provides a robust constraint on the reionization history. There
seems a trouble in the precise definition of the upper limit of the
above integral, because of the present poor knowledge of the onset
of star formation. However, due to the rapid decease of the degree
of ionization with increasing redshifts (see the results presented
in the next section), the CMB optical depth could be mainly
contributed by the electrons at relatively low redshifts $z\ll
z_{h}$. More importantly, since the {\it WMAP} experiment is
probably insensitive to too high redshifts \cite{nolt09,lars11},
the value of $z_h$ actually can not be taken to extremely high.
Therefore, when we solve eq. (\ref{xz}) from high to low redshifts,
we somewhat arbitrarily take the initial redshift $z_h$ to be a
moderate value of 20 at which the transition from Population (Pop)
III to I/II stars\footnote{The Pop III-I/II transition is usually
considered to take place when the cosmic metallicity is enriched to
$\sim 10^{-3.5}Z_{\odot}$. However, the recently discovered Caffau
star showed that the critical metallicity could be much smaller due
to dust cooling, i.e., $\sim10^{-5}Z_{\odot}$ \cite{schn12}. So the
transition could proceed more rapidly than previously considered.}
may take place \cite{furl05,brom06,grei06,tan12}. Correspondingly,
the value of $x$ at $z_h$ is in principle determined by both the
residual electrons left from the recombination era \cite{seag00}
and, in particular, the electrons ionized by possible Pop III stars
at $z>z_h$. For a generally consideration, a wide range of $0\leq
x(z_h)\leq 0.1$ would be taken into account in our calculations.
Here all of the uncertain physics above redshift $z_h$ is packed
into the free parameter $x(z_h)$. Finally, it should be noticed that
the contribution to the CMB optical depth by the residual electrons
was already accounted for when deriving $\tau$ from the CMB data.
Therefore, in principle, the fraction of these electrons needs to be
deducted from the total $x(z)$ when we calculate the optical depth
by eq. (\ref{tau1}). Nevertheless, in view of the actually small
fraction of the residual electrons on the order of $10^{-3}$
\cite{seag00}, such a deduction would be ignored in our
calculations.

\begin{figure}
\centering\resizebox{0.7\textwidth}{!}{\includegraphics{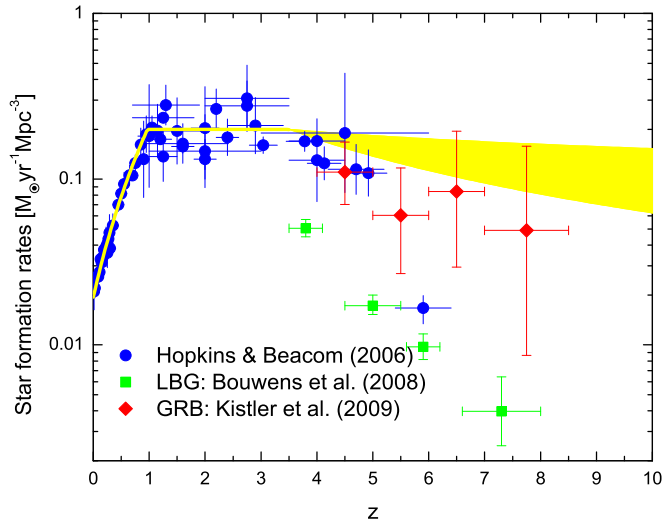}}
\caption{Star formation history with $0.3\leq \alpha\leq1.3$ (shaded
band), which is constrained by both the {\it WMAP-7} CMB optical
depth (i.e., $\tau=0.088$) and the Gunn-Peterson trough observations
(i.e., $6\lesssim z_{\rm full}\lesssim 7$). Some
observationally-determined SFRs are also presented for comparisons.}
\end{figure}

Solving eqs. (\ref{xz}) and (\ref{tau1}) depends on two crucial but
controversial astrophysical inputs, i.e., the star formation rate
$\dot{\rho}_*(z)$ and the clumping factor $C(z)$. Following a series
of measurements of SFRs, especially the complication of Hopkins \&
Beacom \cite{hopk06} which is shown in figure 1 (solid circles), a
consensus on the history of star formation now emerges up to
redshift $z\sim3.5$. The behavior includes a steady increase of star
formation from $z=0$ to $z=1$, and a following plateau up to
$z\sim3.5$, which can be empirically expressed by
\begin{equation}
\dot{\rho}_*(z)\propto\\
\left\{
\begin{array}{lcl}\left({1+z}\right)^{3.44}, & {\rm
for}~
z\leq0.97,\,\\
\left({1+z}\right)^{0}, & {\rm for~} 0.97\leq z\lesssim 3.5,\,
\end{array}\right.
\label{sfr1}
\end{equation}
with a local rate $\dot{\rho}_*(0)=0.02~ M_\odot ~\rm yr^{-1}
Mpc^{-3}$. For $z\gtrsim 3.5$, a clue to a decrease in the SFRs was
seemingly implied by the Hopkins \& Beacom's data, which was further
confirmed by the observations of LBGs and GRBs. Anyway, in view of
the obvious insufficiency of the high-redshift observations, the
tendency of the star formation history to high redshifts is in fact
still ambiguous. Therefore, all of the cases including those in
which the history continues to plateau, drops off, or even increases
were taken into account in previous literature (e.g.,
\cite{daig06}). Instead of a theoretical modeling or observational
determination, we here introduce a free parameter $\alpha$ to
parameterize the high-redshift history as a power law
\begin{equation}
\dot{\rho}_*(z)=0.2\left({1+z\over4.5}\right)^{-\alpha}~{\rm
M_{\odot}~yr^{-1}Gpc^{-3}},~{\rm for~}z\gtrsim3.5.\label{sfr2}
\end{equation}
Such a simple power-law behavior would be regarded as the first
order approximation of the realistic history. Finally, for the
clumping factor, its value can be estimated by numerical simulations
(e.g., \cite{gned97,shul11}) and by semianalytical models (e.g.,
\cite{mada99,chiu00,haim01,wyit03}). All of these calculations
showed that the clumping factor decreases with increasing redshifts,
which can approximately be described by a power law. Following a
recent simulation given by ref. \cite{shul11}, here we adopt
\begin{equation}
C(z)=2.9\left(1+z\over6\right)^{-1.1},~{\rm for}~z\gtrsim
5\label{clumb}
\end{equation}
which has a fixed $C(z<5)=2.9$.

\section{Results}
Substituting eqs. (\ref{sfr1}-\ref{clumb}) into eq. (\ref{xz}), a
cosmic reionization history, $x(z)$, can be solved for a set of
values of the parameters $\mathscr{N}_{\gamma}$ and $\alpha$. By
equating the CMB optical depth for this reionization to the {\it
WMAP-7} one (i.e., $\tau=0.088$), we get a nearly linear
relationship between the parameters $\lg \mathscr{N}_{\gamma}$ and
$\alpha$, as presented in the top panel of figure 2, which can be
fitted by
\begin{eqnarray}
\alpha=a\lg{\mathscr{N}_{\gamma}}-b\label{alpha-ngf}
\end{eqnarray}
with $a=2.18\pm0.05$ and $b=3.89\pm0.01$. Here the error bars of the
coefficients are resulted from the uncertainty of the parameter
$x(z_h)$. Moreover, the small values of the error bars indicate that
the $\mathscr{N}_{\gamma}-\alpha$ relationship is actually
insensitive to the value of $x(z_{h})$, which demonstrates that it
is acceptable to ignore the deduction of the fraction of residual
electrons. With the above relationship, we present the solutions to
eq. (\ref{xz}) in figure 3 for different values of $\alpha$, where
the redshift at which the IGM is fully ionized (denoted by $z_f$)
can be found. Subsequently, we can obtain the relationship between
$z_f$ and $\mathscr{N}_{\gamma}$, as shown in the bottom panel of
figure 2.
\begin{figure}
\centering\resizebox{0.5\textwidth}{!}{\includegraphics{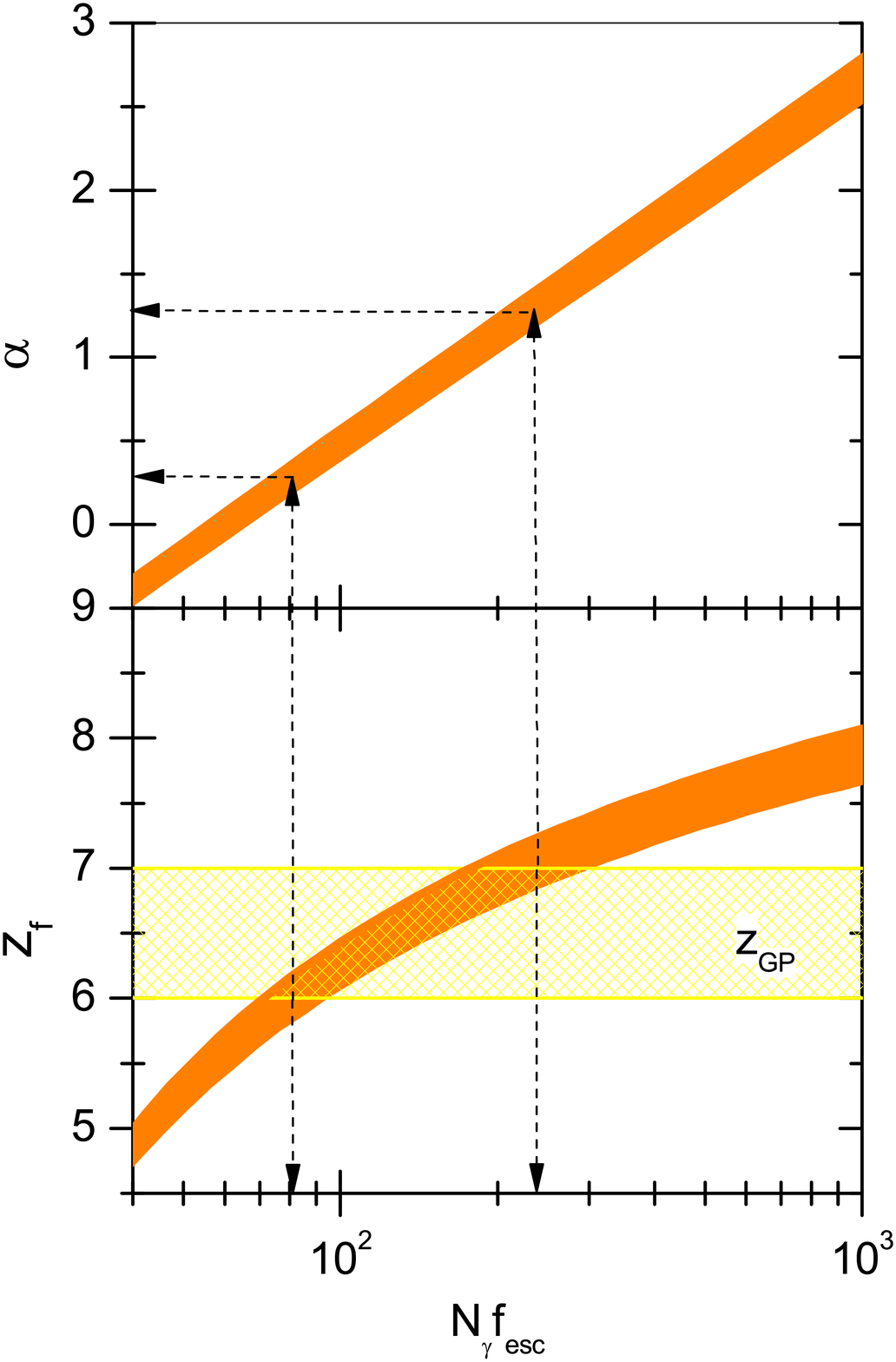}}
\caption{Relationships required by the {\it WMAP-7} CMB optical
depth ($\tau=0.088$) between the parameters $\mathscr{N}_{\gamma}$
and $\alpha$ (top panel) or $\mathscr{N}_{\gamma}$ and $z_{\rm
full}$ (bottom panel), as shown by the solid bands. The width of the
bands is determined by the adopted range of the initial degree of
ionization as $0\leq x(z_h)\leq0.1$. The possible range of $z_f $
denoted by the hatched band is inferred from the Gunn-Peterson
trough observations, and consequently the dashed arrows represent
the constraints on the parameters of $\alpha$ and
$\mathscr{N}_{\gamma}$.}
\end{figure}

As also shown by figure 3, the fittings to the reionization
histories for $z_{f}<z \lesssim10$ exhibit a power-law behavior as
\begin{equation}
x\approx\left({1+z\over 1+z_{f}}\right)^{-(\alpha+{1.7})},~~{\rm
for}~z\geq z_{f}.\label{xapp}
\end{equation}
Substituting the above approximation into eq. (\ref{tau1}), an
analytical expression for the optical depth can be obtained
\begin{eqnarray}
\tau&\approx&{2\sigma_Tn_H^0c(1+y)\over
3H_0\Omega_m^{1/2}}\left[(1+z_{f})^{3/2}\left(1+{15\eta\over2+10\alpha}\right)-{1\over\Omega_m^{1/2}}\right],
\end{eqnarray}
where $\eta=1-[(1+z_h)/(1+z_f )]^{-\alpha-1/5}$ and the parameter
$\Omega_\Lambda$ is omitted due to $\Omega_\Lambda\ll \Omega_m(1+z_f
)^3$. For the {\it WMAP-7} optical depth $\tau=0.088$, the redshift
for full reionization as a function of $\alpha$ can be expressed
\begin{eqnarray}
z_{f}\approx11.7\left({1+{15\eta\over2+10\alpha}}\right)^{-2/3}-1.\label{zre-alpha}
\end{eqnarray}
Obviously, for $\alpha\rightarrow \infty$ and thus $\eta\rightarrow
1$ which is the case considered in the {\it WMAP} analyses, the
value of $z_f $ approaches to its upper limit 10.7. In contrast, for
a more realistic value of $\alpha$ on the order of unity, eq.
(\ref{zre-alpha}) clearly shows that the redshift for full
reionization could be much lower than its upper limit.

\begin{figure}
\centering\resizebox{0.7\textwidth}{!}{\includegraphics{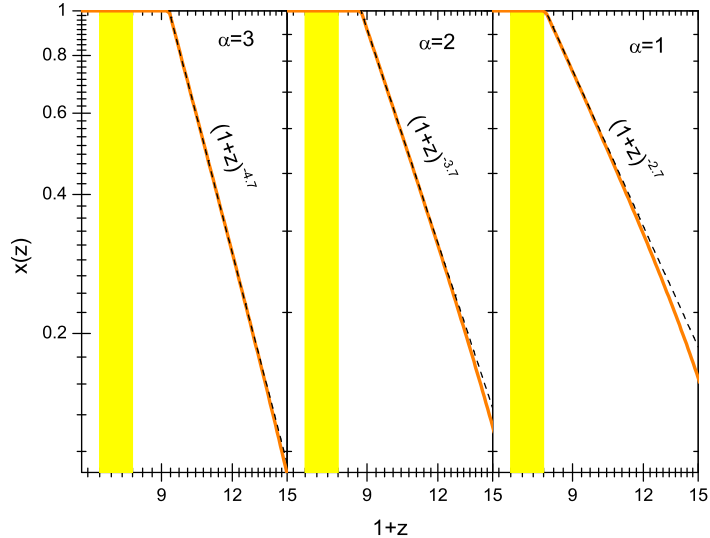}}
\caption{Numerically calculated evolutions of the ionization degree
of hydrogen with redshift for different $\alpha$ (solid lines),
where the values of $\mathscr{N}_{\gamma}$ are taken according to
eq. (\ref{alpha-ngf}) with $x(z_h)=0$. The dashed lines represent
power-law fittings to the numerical results.}
\end{figure}

\section{Discussions}
The results presented above demonstrate that the effective power-law
approximation for the high-redshift star formation history can
dramatically simplify the calculations of reionization and CMB
optical depth. More usefully, the connections among the star
formation, reionization, and CMB optical depth could be
approximately analytically addressed. These results make it
convenient to test the theoretical models of star formation and
reionization quickly and effectively. At the same time, our results
also provide an observational base to some relevant astrophysical
studies, e.g., the estimation of the event rate of high-redshift
GRBs.

As implied by the probes to the Gunn-Peterson trough (Ly$\alpha$
absorption) toward high-redshift quasars and galaxies, the full
reionization of hydrogens could take place not far beyond $z_{\rm
GP}\sim6$ \cite{beck01,fan02,fan06,gned06}. Of course, by
considering that the Gunn-Peterson test is actually sensitive to
small neutral fractions, the value of $z_f $ could be somewhat
higher. For example, detections of Ly$\alpha$-emitting galaxies at
$z=6.6$ \cite{hu02,koda03} suggested that the redshift for full
reionization might be as high as $z\approx7$. Therefore, for an
independent and supplementary consideration, if we take $6\lesssim
z_f \lesssim7$, a constraint on the star formation history as well
as on the parameter $\mathscr{N}_{\gamma}$ could be derived from the
$\mathscr{N}_{\gamma}-\alpha$ and $\alpha-z_{f}$ relationships:
\begin{eqnarray}
0.3\pm0.2\lesssim\alpha\lesssim1.3\pm0.3,\label{alpha range}
\end{eqnarray}
and
\begin{eqnarray}
80\pm12\lesssim\mathscr{N}_{\gamma}\lesssim230\pm60.
\end{eqnarray}
On one hand, the range of $\alpha$ is well consistent with the one
constrained by the observed near infrared excess in the cosmic
infrared background spectrum \cite{tan12}. On the other hand, for a
typical value of 4000 of $N_{\gamma}$ for normal stars, the range of
$\mathscr{N}_{\gamma}$ indicates that the escaping fraction $f_{\rm
esc}$ is within the range from $2.0\pm0.3\%$ to $5.8\pm1.5\%$. This
is well consistent with the previous observations and calculations.

The possible star formation history with
$0.3\lesssim\alpha\lesssim1.3$ (shaded band) is presented in figure
1 in comparison with some observations. Then the comparison shows
that, in general, the SFRs inferred from both GRBs and LBGs are too
low to reionize the universe. To be specific, the slight shortage of
the GRB-inferred SFRs may suggest that the calibration between the
GRBs and stars needs to be investigated more carefully. For example,
some selection effects due to redshift measurements \cite{cao11}
should be taken into account, which are missed by ref.
\cite{kist09}. On the other hand, as also argued by some previous
works (e.g., \cite{kist09,tren12,chou07}), the significant
discrepancy between the LBG-inferred SFRs and the required levels
indicates that most of the high-redshift star formation could take
place in very small halos hosting dwarf galaxies, which are too
faint to be observed in deep LBG surveys.

The result present above robustly exhibits the importance and
advantage of the GRB observations in the SFR determination. In some
previous works (e.g., \cite{chen10}), the SFRs inferred from
high-redshift GRBs were considered to be too high in contrast to the
relatively low LBG-inferred SFRs. Therefore, Cheng et al.
\cite{chen10} proposed that these GRB-inferred SFRs may be
overestimated due to that some high-redshift GRBs do not originate
from the collapse of massive stars, but could be produced by the
bursts of superconducting cosmic strings. However, the result
addressed in this paper indicates an opposite situation, i.e., that
the GRB-inferred SFRs actually are not overestimated but somewhat
underestimated. Therefore, the expected number of the bursts of
superconducting cosmic strings could be much lower than the one
estimated in ref. \cite{chen10}.

\begin{figure}
\centering\resizebox{0.5\textwidth}{!}{\includegraphics{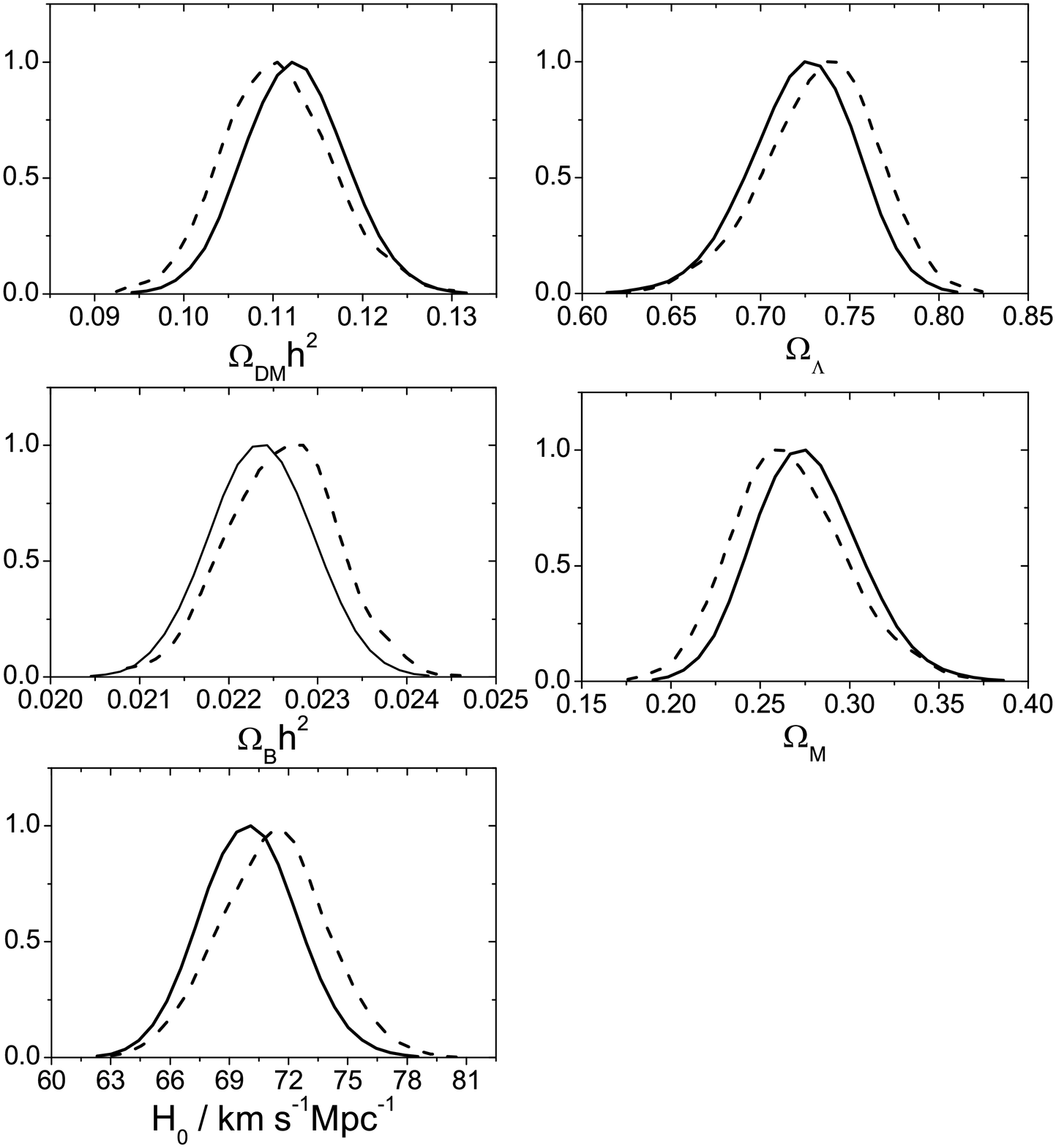}}
\caption{The probability distributions of some cosmological
parameters resulting from a fitting to the CMB power spectra with a
hyperbolic-tangent-function reionization (dashed lines) or a
power-law reionization as eq. (\ref{xapp}) (solid lines).}
\end{figure}

Finally, it is worth noticing that the values of the cosmological
parameters we adopted above actually are obtained by fitting the CMB
power spectra with a hyperbolic-tangent-function reionization.
Therefore, as pointed out by some previous works (e.g.,
\cite{pand11,mitr12}), it seems necessary to revisit the fitting to
the CMB power spectra with a more realistic reionization history,
e.g., the power-law behavior as presented in eq. (\ref{xapp}). For a
preliminary attempt, with the CosmoMC code \cite{pett08} (updated
CAMB), we plot the probability distributions of several cosmological
parameters in figure 4, both for the cases of tangent-function and
power-law reionizations. In comparison, the modification of the
parameters due to the new reionization module is basically apparent
and, specifically, the peak values of the parameters shift about few
percent. Nevertheless, such shifts would not significantly influence
the results obtained in this paper.

\acknowledgments The authors acknowledge the anonymous referee for
his/her helpful comments. YWY is supported by the National Natural
Science Foundation of China (Grant Nos. 11103004 and 11073008) and
by the Self-Determined Research Funds of CCNU from the colleges¡¯
basic research and operation of MOE of China (Grant Nos.
CCNU12A01010). KSC is supported by the GRF Grants of the Government
of the Hong Kong SAR under HKU7010/12P. MCC is supported by grants
from the Research Grant Council of the Hong Kong SAR (Project No .
400910).

\end{document}